\documentclass[10pt,twocolumn]{article}
\usepackage[draft]{hyperref}
\usepackage{amsmath}
\usepackage{amssymb}
\usepackage{graphicx}
\usepackage{stfloats}
\usepackage{placeins}
\usepackage{fullpage}
\usepackage{fixltx2e}
\usepackage{soul}
\usepackage[articletitle=true,maxauthors=20]{achemso}

    \makeatletter
    \let\@fnsymbol\@arabic %
    \makeatother
    
\begin{document}

\twocolumn[

\title{Resonant Electro-Optic Imaging for Microscopy at\\ Nanosecond Resolution}

\author{Adam J. Bowman\textsuperscript{*} and
            Mark A. Kasevich
           \\ \footnotesize{ Physics Department, Stanford University, 382 Via Pueblo Mall, Stanford, California 94305, USA} 
           \\\textsuperscript{*}\footnotesize{abowman2@stanford.edu}}
\date{}
\maketitle
\subsection*{Abstract}
\noindent\textbf{We demonstrate an electro-optic wide-field method to enable fluorescence lifetime microscopy (FLIM) with high throughput and single-molecule sensitivity. Resonantly driven Pockels cells are used to efficiently gate images at 39 MHz, allowing fluorescence lifetime to be captured on standard camera sensors. Lifetime imaging of single molecules is enabled in wide-field with exposure times of less than \textbf{100} milliseconds. This capability allows combination of wide-field FLIM with single-molecule super-resolution localization microscopy. Fast single-molecule dynamics such as FRET and molecular binding events are captured from wide-field images without prior spatial knowledge. A lifetime sensitivity of 1.9 times the photon shot-noise limit is achieved, and high throughput is shown by acquiring wide-field FLIM images with millisecond exposure and  $>10^8$ photons per frame. Resonant electro-optic FLIM allows lifetime contrast in any wide-field microscopy method.\\\\
}

\vspace{3mm}
]

\begin{figure}[t!]
  \includegraphics[width=\columnwidth]{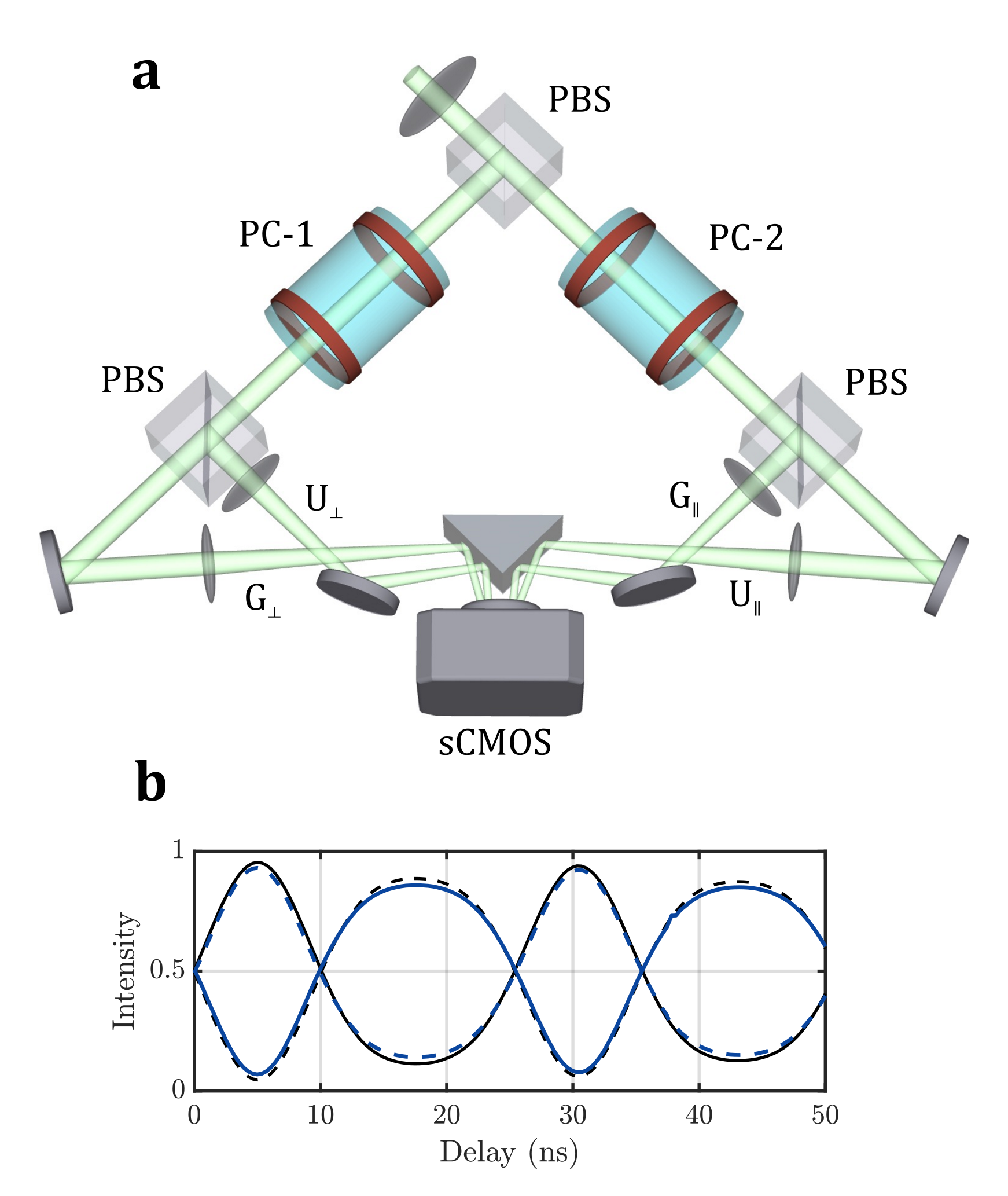} 
\caption{\label{fig:1}  {\footnotesize \textbf{Resonant EO-FLIM.} (a) Schematic of dual PC resonant EO-FLIM detection. Fluorescence emission is split on a first polarizing beamsplitter (PBS) and each polarization component is then modulated by a separate PC. A second PBS splits gated (G) and ungated (U) image intensities from the modulated beams, which are determined by Eqs. 1 and 2. The four images are then combined on a single sCMOS camera. (b) Gating functions are measured for each output image for the parallel (blue line/dash) and perpendicular (black line/dash) polarization channels by varying time delay (or equivalently, phase shift) between excitation and PC drive while imaging laser reflection. 
}}
\end{figure}

\begin{figure*}
\centering
  \includegraphics[width=0.8\textwidth]{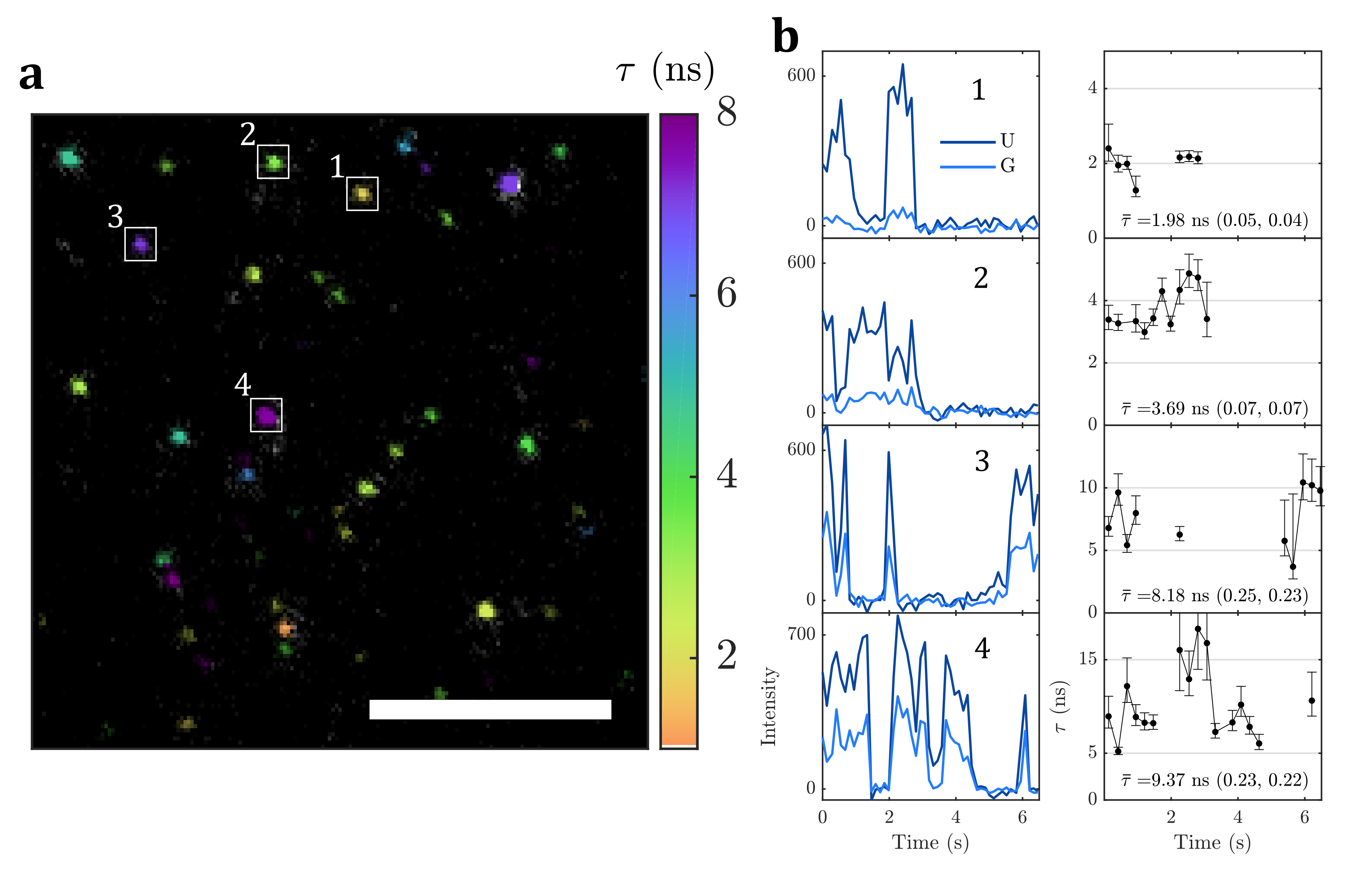}
  \vspace{-5.0mm}
\caption{\label{fig:4} {\footnotesize \textbf{Wide-field single-molecule FLIM.} (a) A mixture of Alexa 532, Alexa 546, and CdSe/ZnS quantum dot (QD) single emitters in a spin coated PMMA film (scale bar 10 $\mu$m). The image results from the combination of ten 100 ms exposures. (b) Single-molecule intensity and lifetime trajectories are acquired in wide field for the indicated molecules in (a). Plots 1-4 for the corresponding molecules illustrate how the intensity ratio from the ungated (U) and gated (G) images maps to different lifetime estimates. Average lifetime measurements over the entire trace are also given for each molecule. Emitter 4 (likely a quantum dot) shows a clear jump in lifetime. Molecule population distributions are also characterized in Supplementary Fig. 5. 
}}
\end{figure*}

\noindent
Improving the chemical resolution of imaging measurements is an outstanding challenge in microscopy. Fluorescence lifetime is a promising contrast mechanism \cite{Berezin2010FluorescenceImaging,Datta2020FluorescenceApplications}, but existing approaches sacrifice throughput and efficiency. Single-molecule techniques such as super-resolution microscopy and FRET particularly stand to benefit from lifetime contrast, but they are difficult with existing detectors \cite{Oleksiievets2020Wide-FieldMolecules,Thiele2020ConfocalMicroscopy}.  We present a method that uses resonantly-driven optical modulators to obtain sub-nanosecond resolution on standard cameras with performance near the photon shot-noise limit. Resonant electro-optic fluorescence lifetime microscopy (EO-FLIM) enables single-molecule imaging in wide-field at acquisition rates greater than 1000 times the existing state-of-the-art, and it allows nanosecond wide-field imaging in high sensitivity applications. For example, improved chemical resolution is desirable to unmix the signatures of fluorescent probes for multiple labeling \cite{Niehorster2016a,Valm2016MultiplexedLabels,Valm2017}, to improve measurements of bio-molecular dynamics and interactions \cite{Eggeling2001DataDetection,Prummer2004MultiparameterAnalytics,Watkins2004InformationMeasurements,Esposito2019EnhancingMicroscopy,Trinh2019FastApplications}, and to enable label-free chemically-specific contrast \cite{Cheng2015VibrationalMedicine,You2018IntravitalMicroscopy}.  These goals require moving beyond standard spatial and spectral imaging channels. 

In FLIM microscopy, short ($\sim 100$ ps) optical pulses are used to excite molecules, which subsequently emit fluorescence photons on 100 picosecond to 10 nanosecond timescales. The radiative lifetime provides a sensitive probe of local physical and chemical parameters as well as the available decay pathways. Since single molecules emit at most one photon per pulse, MHz excitation optimizes the rate at which fluorescent photons are imaged.  Resonant EO-FLIM combines high excitation rates (39 MHz) with single-molecule sensitivity. While there have been several approaches developed for wide-field FLIM, they fail to achieve the performance required for single molecule imaging. SPAD arrays are a promising and rapidly developing technology for FLIM, \cite{Gyongy2017256256Applications,Gyongy2016Smart-aggregationCameras,Antolovic2017SPADBlinking, Ulku2018AFLIM,Bruschini2019Single-photonOutlook,Henderson2019ATechnology,Zickus2020FluorescenceEstimation, Nedbal2021CorrectionImaging,Sagoo2021RapidApplications} but they still face obstacles. Large pixel arrays can achieve time-gated imaging \cite{Gyongy2017256256Applications, Zickus2020FluorescenceEstimation} but are not yet competitive with scientific camera sensors. Realizing on-chip timing electronics remains a challenge\cite{Bruschini2019Single-photonOutlook}, and SPADs are also subject to fundamental pile-up effects due to pixel dead time. Other detectors like gated optical intensifiers  \cite{Sparks2017} and modulated CCD and CMOS sensors \cite{Mitchell2002DirectImaging,Esposito2005All-solid-stateSensing,Franke2015Frequency-domainSensor,Raspe} are noisy and inefficient, falling significantly short of the pixel performance of standard scientific cameras. Gated intensifiers have high photon loss due to the photocathode quantum efficiency and because they only capture the gated light fraction. They also suffer from gain noise and reduced image quality unless operated in low-throughput photon counting mode \cite{Hirvonen2015Single-moleculeImaging}.  Similarly, modulated pixels suffer from high read noise and dark current. 

Confocal scanning with time-correlated single photon counting detectors remains the standard method for FLIM microscopy \cite{Becker1999Time-resolvedTCSPC}. These detectors are limited to MHz count rates for a single scanned pixel. To understand the limitations of confocal scanning, consider a single molecule image composed of $10^4$ diffraction-limited pixels. To image at a signal level of 1000 photons per molecule with a typical 2 MHz count rate, a scanning acquisition over all pixels requires 5 seconds. Wide-field imaging removes this temporal bottleneck. In this example, all $10^4$ pixels may instead be acquired simultaneously, in principle allowing 4 orders of magnitude faster acquisition on single-molecule samples. For bright samples having more than one molecule per pixel, TC-SPC becomes limited by photon pile-up due to the dead time of the counting detector. Resonant EO-FLIM instead allows multiple photons to be detected per pulse at each pixel while using optimal excitation rates, theoretically allowing 6-8 orders of magnitude throughput improvement for densities of $\sim$10-1000 molecules per pixel at saturation. EO-FLIM allows photon-efficient lifetime estimation in wide-field from a single camera exposure --- all images reported here result from a single frame with no multi-point sampling or fitting common to other FLIM approaches. 
 
Prior demonstrations of single-molecule FLIM have significant limitations. Our proof-of-concept at kHz repetition rates \cite{Bowman2019Electro-OpticMicroscopy} allowed wide-field FLIM imaging of single-molecules only at long exposures. A hybrid image intensifier-based TC-SPC detector\cite{Hirvonen2017} at $<$1 MHz count rate has also been demonstrated\cite{Oleksiievets2020Wide-FieldMolecules}, but it suffers from low throughput and low photocathode quantum efficiency, requiring minutes per acquisition.  Super-resolution FLIM has previously been achieved by confocal STED \cite{Lesoine2012SupercontinuumImaging,Niehorster2016a}, confocal image scanning microscopy \cite{Castello2019AFLIM}, and recently in single-molecule localization by using fast confocal scanning over a limited field of view \cite{Thiele2020ConfocalMicroscopy}. However, there has been no prior technique sensitive enough to enable super-resolved FLIM acquisition in wide-field. Compared to confocal scanning, resonant EO-FLIM allows parallel observation of all image pixels, which can theoretically improve throughput and imaging speeds by several orders of magnitude. Our single-molecule demonstrations are currently limited by laser power, but higher power sources will allow faster single-molecule imaging speeds in the future. The physical limit on acquisition speed corresponds to excitation of each image pixel at single-molecule saturation (approximately 10-100 $\mu$W per diffraction-limited region)\cite{Michalet2002Single-moleculeMicroscopyb,Lin2015QuantifyingSpeeds}.

We demonstrate the capabilities of resonant EO-FLIM for single-molecule imaging. Molecular populations are imaged in wide-field with exposures less than 100 ms, FLIM is combined with super-resolution localization microscopy, and rapid single-molecule dynamics including FRET and binding events are imaged.

\noindent
\subsection*{Results and Discussion}
\noindent
\textbf{Resonant EO-FLIM.} Pockels cells (PCs) are used to capture fluorescence lifetimes by intensity modulation. By gating wide-field images all-optically, they allow nanosecond lifetimes to be captured in static frames that are acquired on a standard sCMOS camera \cite{Bowman2019Electro-OpticMicroscopy}. The PC acts as a voltage-controlled waveplate that applies a time-dependent birefringent phase shift to initially polarized imaging beams as shown in Fig. \ref{fig:1}(a). The birefringent phase shift is then converted to intensity modulation by second polarizing beamsplitters in each initial polarization path. This allows all of the input light to be recorded in four spatially separated images on the camera sensor while simultaneously encoding temporal information.

A resonant drive is applied to 10 mm aperture KD*P Pockels cells that have been modified to allow for liquid cooling. A mode-locked supercontinuum fiber laser (NKT Photonics EXW-12) is synchronized with the high-voltage drive at a pulse repetition rate of 39 MHz. The Pockels cell acts as the capacitor in an radio frequency (RF) tank circuit to resonantly enhance drive voltage (Supplementary Fig. 1). As a result, it is possible to create optical modulation of an image with 86$\%$ modulation depth and 16.2 dB extinction (Supplementary Fig. 2). This modulation is recorded over one drive period in Supplementary Video 1. The gating function realized is not sinusoidal but takes the form $g(t) = \sin^2[A\pi\sin(\omega t)/2]$  where $A$ corresponds to the high-voltage drive strength in units of $V_\pi$. The gating function is periodic with twice the frequency of the RF drive at $\omega$. Increasing $A$ leads to a sharper modulation function due to harmonic content. For example in Fig. \ref{fig:1}(b), a gating slope $13\%$  steeper than an equivalent sinusoidal gate is achieved. The sample's fluorescence intensity $I(t,\tau)$ is convolved with the gating function, giving gated (G) and ungated (U) image intensities after the second beamsplitter determined by
\begin{align}
    G(\phi_0,\tau) &= \int I(t,\tau)g(t-\phi_0/\omega) \mathrm{d}t\\
U(\phi_0,\tau) &= \int I(t,\tau)[1-g(t-\phi_0/\omega)]\mathrm{d}t. 
\end{align} Here $\tau$ is the unknown lifetime parameter being estimated, but more generally it could be any unknown temporal parameter of the image. G and U are shown in Fig. \ref{fig:1}, with the G beam path corresponding to crossed polarizers. By changing the relative phase $\phi_0$ between the laser pulse train and Pockels cell drive, the gated and ungated convolutions are directly measured and may be fit or deconvolved in order to determine the fluorescence decay function $I(t,\tau)$. Further, because polarization information is captured along with lifetime, EO-FLIM may be used to measure time-resolved polarization anisotropy. This is a valuable probe of the rotational correlation time of molecules in solution  \cite{Siegel2003Wide-fieldFluorophore,Smith2015AData} and may also be used to detect homo-FRET \cite{Teijeiro-Gonzalez2021Time-ResolvedDimer}. Example time traces measuring fluorescence lifetime and time-resolved anisotropy are shown in Supplementary Fig. 3.

When estimating a single-exponential decay, the ratio of the gated and ungated output intensities G/U at a single phase may be used to estimate lifetime from a single camera frame. This allows real-time observation of fluorescence lifetime at the camera frame rate. Single-frame estimation allows observation of single-molecules and dynamic samples where fluorescence intensities change rapidly from frame to frame. A lookup table is generated numerically from the measured gating function to map each measured intensity ratio to a corresponding fluorescence lifetime (Supplementary Fig. 4).  Lifetime uncertainties are primarily set by the number of collected photons and the RF modulation frequency, as described below. This single-frame method is used for all wide-field images reported below.\\

\noindent

\begin{figure*}[ht!]
\centering
  \includegraphics[width=\textwidth]{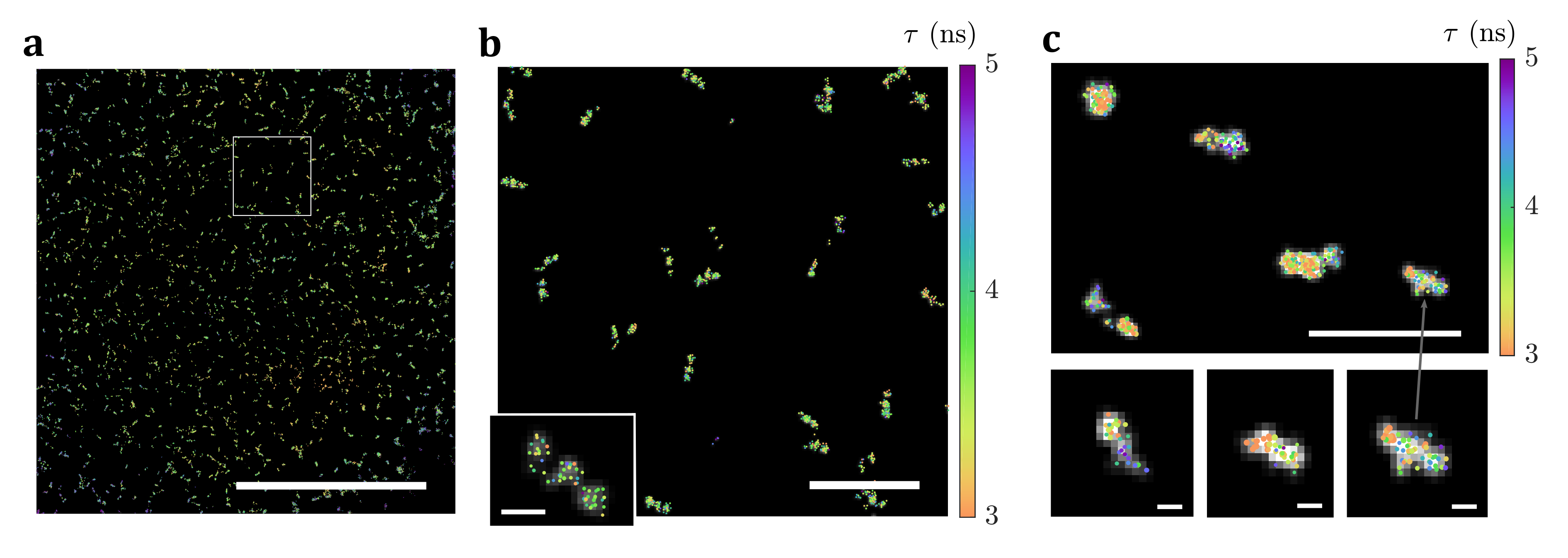}
  \vspace{-5.0mm}
\caption{\label{fig:5} {\footnotesize \textbf{Wide-field super-resolution FLIM.} (a) DNA-PAINT is performed on DNA origami structures using Atto 542 (Gattaquant 80G) in TIRF microscopy. Each single-molecule localization is associated with a lifetime estimate (Supplementary Video 2). Lifetime for each localization is represented by a colored dot overlaid on the super-resolution images. A wide field of view is shown resulting from 1000 acquired frames at 150 ms exposure (scale bar 10 $\mu$m). (b) A magnified view of the ROI indicated in (a) demonstrates clearly resolved binding sites separated by 80 nm (scale bar 1 $\mu$m, inset 100 nm). This sample yields $\sim$4000 photons per localization. (c) Dual-dye DNA origami structures (Cy3B and Atto 565) with alternating 80-80-20 nm spacings are imaged. Due to the filter set used this sample gave $< 800$ photons per localization and single sites are not spatially resolved, but lifetime differences in binding sites along the structures are able to be seen (scale bar 1 $\mu$m, insets 100 nm). Population distributions are shown in Supplementary Fig. 6. 
}}
\end{figure*}

\begin{figure*}[ht!]
\centering
  \includegraphics[width=0.9\textwidth]{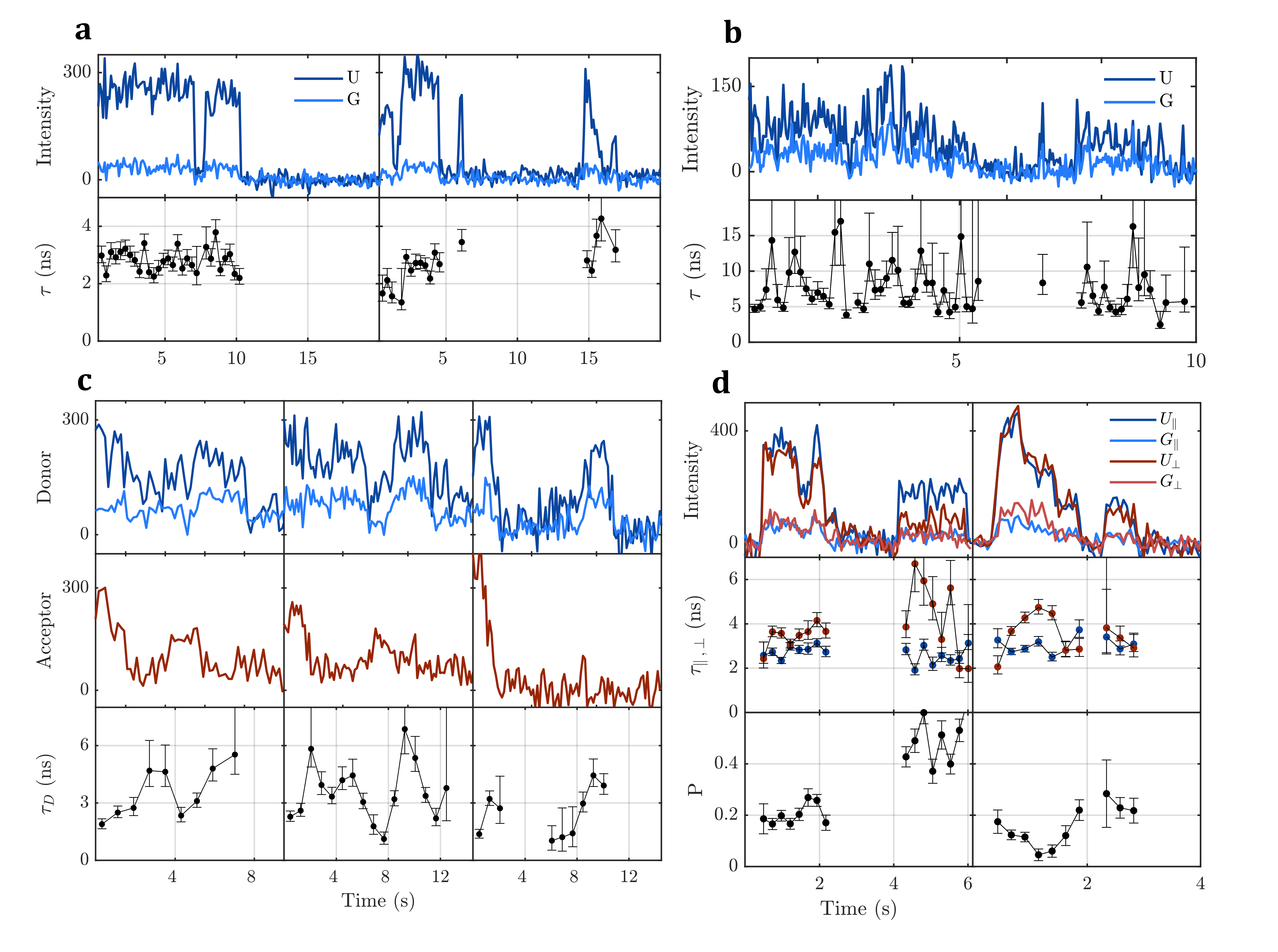}
  \vspace{-0.0mm}
\caption{\label{fig:6} {\footnotesize \textbf{Single-molecule lifetime dynamics.}  Example traces of single-molecule intensity, lifetime, and multi-parameter measurements are shown. (a) Two Alexa 532 molecules in PMMA are plotted, with the second showing a jump in lifetime (9 Hz frame rate, 3 Hz lifetime estimates). (b) A single CdSe/ZnS quantum dot shows rapid intensity and lifetime fluctuations (33 Hz frame rate, 8.3 Hz lifetime estimates) (c) Single-molecule FLIM-FRET is performed using Atto 532 (donor) and Atto 647N (acceptor) on fixed DNA origami structures with 4.1 nm separation. Three molecules are shown. Lifetime for the donor is observed to decrease during periods of acceptor emission, indicating FRET quenching of the donor (6.3 Hz frame rate, 1.3 Hz lifetime estimates). (d) Multi-parameter measurement of single-molecule DNA-PAINT bursts. Intensity, lifetime (parallel and perpendicular polarization channels), and static polarization are plotted. The perpendicular lifetime estimate is larger than the parallel estimate, indicative of time-resolved anisotropy associated with the rotational motion of the fluorophore (17 Hz frame rate, 4.3 Hz lifetime estimates). Due to the binning of several intensity frames for each lifetime estimate, lifetime transitions such as in the FRET traces in (c) may be smoothed, and estimates at burst edges in (a, d) may have fewer photons and increased error.
}}
\end{figure*}

\noindent
\textbf{Wide-field single-molecule FLIM.} Imaging single-molecules in wide-field enables applications in super-resolution microscopy \cite{Liu2015} and FRET \cite{Lerner2021ietPractices}, particularly when high throughput is desired. To characterize resonant EO-FLIM's capability for single-molecule imaging, we first demonstrate FLIM of static single molecules in spin-coated films of PMMA as a benchmark. A mixture of Alexa 532, Alexa 545, and CdSe/ZnS quantum dots is imaged in Fig. \ref{fig:4}. Lifetime may be estimated for each frame in a time series, allowing dynamic observations of single-emitter trajectories in intensity, macro-time, 2D space, and nanosecond lifetime. This is a substantial improvement over the state of the art for wide-field FLIM, where previous single-molecule demonstrations require tens of seconds \cite{Bowman2019Electro-OpticMicroscopy} or several minutes \cite{Oleksiievets2020Wide-FieldMolecules} of acquisition time. Here we use 200 ms per lifetime estimate. Short exposure times enable observation of single molecule dynamics. Further, by imaging samples having only one molecule type in Supplementary Fig. 5, the single-molecule lifetime distributions can be studied. PMMA films are known to give broad molecular lifetime distributions, consistent with our observations \cite{Macklin1996ImagingInterface,Garcia-Parajo2001OpticalProteins}. The quantum dots show intensity blinking and large lifetime variations [Fig. \ref{fig:4} and Fig. \ref{fig:6}(b)].\\

\noindent
\textbf{Wide-field super-resolution FLIM.} Single molecule FLIM at high frame rate is enabling for super-resolution localization microscopy, where fast molecular bursts are localized for imaging below the diffraction limit. DNA origami samples are imaged using DNA points accumulation in nanoscale topography (DNA-PAINT) with wide-field TIRF excitation to allow super-resolution imaging in Fig. \ref{fig:5}. Supplementary Video 2 demonstrates wide-field lifetime imaging of the single molecule bursts with 150 ms exposure time. This acquisition rate is limited by our excitation intensity of $\sim 0.2$ kW/cm$^2$. DNA-origami structures with three binding sites separated by 80 nm are clearly resolved using Atto 542 fluorophores (Gattaquant PAINT 80G). For each single-molecule localization, we can assign a fluorescence lifetime estimate to allow combination of wide-field FLIM with super-resolution localization microscopy. The intensity and lifetime distribution of the single-molecule localizations may also be studied (Supplementary Fig. 6). In Fig. \ref{fig:5}(c), we can resolve spatial differences in lifetime on a multi-dye sample with separate binding sites for Cy3B and Atto 565 dyes. Because we used a 500-532 nm excitation band for all samples, the more yellow dyes Cy3B and Atto 565 were not as bright as Atto 542 and gave low photon numbers ($< 800$) per localization. \\
\noindent
\noindent

\noindent
\textbf{Rapid FLIM of single-molecule dynamics.} Previous techniques used to study single molecule dynamics with FLIM required spatial knowledge of a molecule's location in order to observe fast dynamics with a confocal detector \cite{Eggeling2001DataDetection,Prummer2004MultiparameterAnalytics,Sorokina2009FluorescentInitiation,Squires2017DirectC-phycocyanin}. EO-FLIM allows wide-field observations without any prior knowledge of a molecule's position. In Fig. \ref{fig:6}, we capture individual intensity and lifetime traces demonstrating several types of single-molecule dynamics. Organic dye molecules like Alexa 532 are relatively stable in lifetime but occasionally show a lifetime jump such as in Fig. \ref{fig:6}(a, right). Quantum dots show a much longer lifetime and rapid intensity and lifetime fluctuations that we can observe with a 33 Hz camera frame rate [Fig. \ref{fig:6}(b)]. By monitoring the quenching of donor fluorophore emission in FRET, it is possible to detect and quantify FRET efficiency without relying on spectral channels and intensity measurements. In Fig. \ref{fig:6}(c), we perform simultaneous lifetime and dual-color FRET detection. Donor lifetime is monitored in one EO-FLIM channel while acceptor intensity is imaged in a separate channel. Reduction in donor lifetime is observed during high-FRET emission bursts from the acceptor. Donor and acceptor molecules are fixed to DNA origami at 4.1 nm separation, so the FRET dynamics here likely result from molecular photophysics such as blinking of the acceptor rather than dynamic changes in pair distance.

We demonstrate that lifetime and polarization information can be combined for measuring rapid single-molecule bursts in DNA-PAINT in Fig. \ref{fig:6}(d). Time-resolved polarization may be captured along with fluorescence lifetime. This is accomplished by using polarized excitation light and comparing the intensities in the parallel and perpendicular EO-FLIM paths defined by the first beamsplitter in Fig. \ref{fig:1} (see also Supplementary Fig. 3). Taking single frames at 50 ms exposure allows individual PAINT binding events to be captured. Since the fluorophores are tethered in solution to DNA origami (with a 4 nucleotide spacer), a polarization anisotropy signal is expected \cite{Kumke1995HybridizationEnhancement}. A static polarization anisotropy associated with the bursts is observed, and the lifetime estimates of the parallel and perpendicular EO-FLIM channels also differ with the parallel channel giving a shorter lifetime. This difference is a signature of time-resolved polarization anisotropy associated with the rotational motion of the fluorophore. Multi-parameter measurements combining intensity, macro-time, lifetime, polarization, and even wavelength appear to be a promising direction for studying single-molecule dynamics \cite{Eggeling2001DataDetection,Prummer2004MultiparameterAnalytics, Squires2017DirectC-phycocyanin}.\\

\begin{figure}[t!]
  \includegraphics[width=\columnwidth]{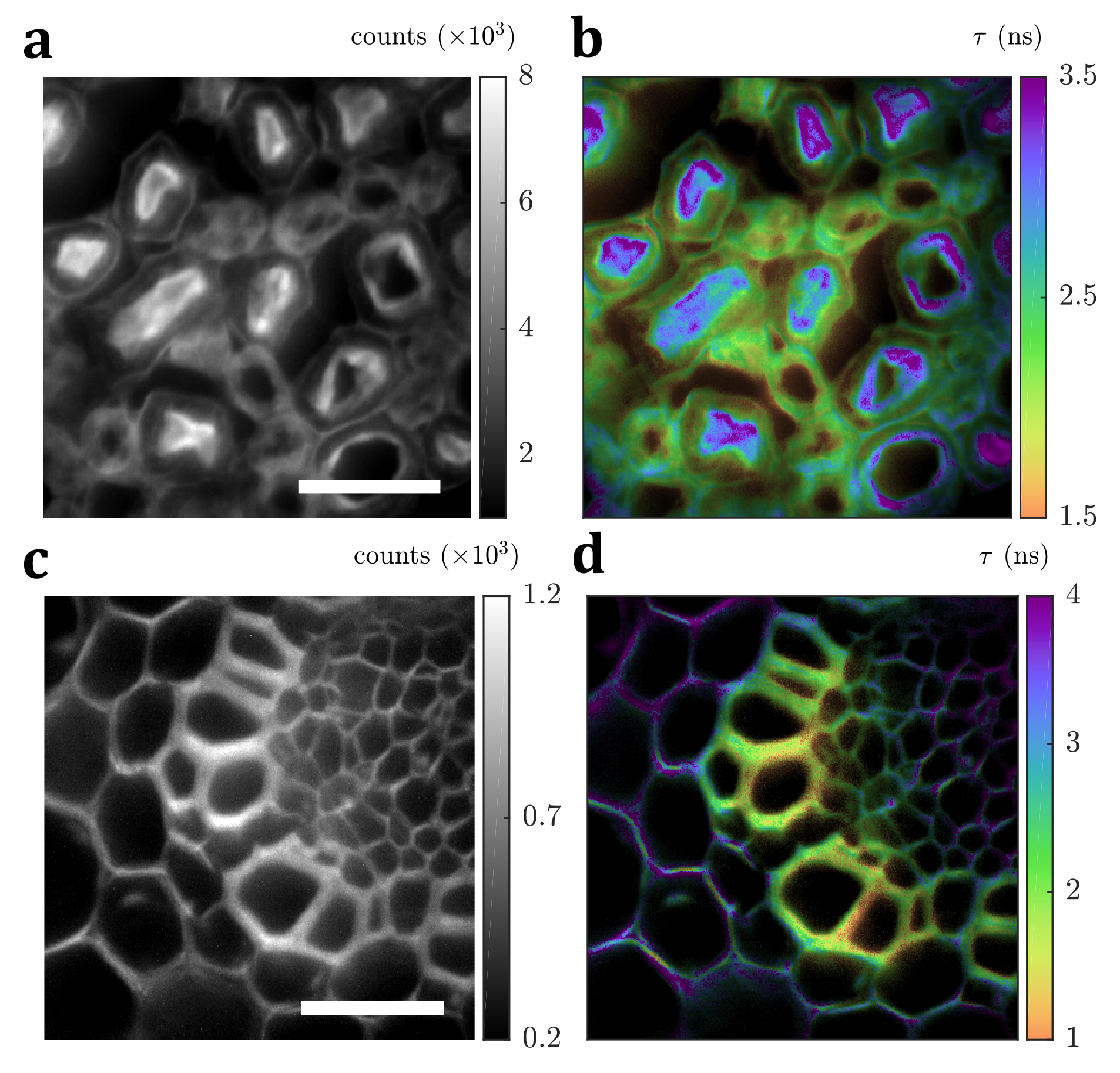}
\caption{\label{fig:3} {\footnotesize \textbf{Rapid wide-field FLIM.} (a) Intensity and (b) lifetime images of a mouse kidney slice stained with Alexa 488 (short lifetime) and Alexa 568 (long lifetime). The dyes are differentiated by lifetime within one spectral channel using a 20 ms exposure. (c) Intensity and (d) lifetime images of a \textit{Convallaria majalis} FLIM sample acquired with $1.2\times 10^8$ photons detected in a 1 millisecond exposure (scale bars 50 $\mu$m).
}}
\end{figure}

\begin{figure}[t!]
  \includegraphics[width=\columnwidth]{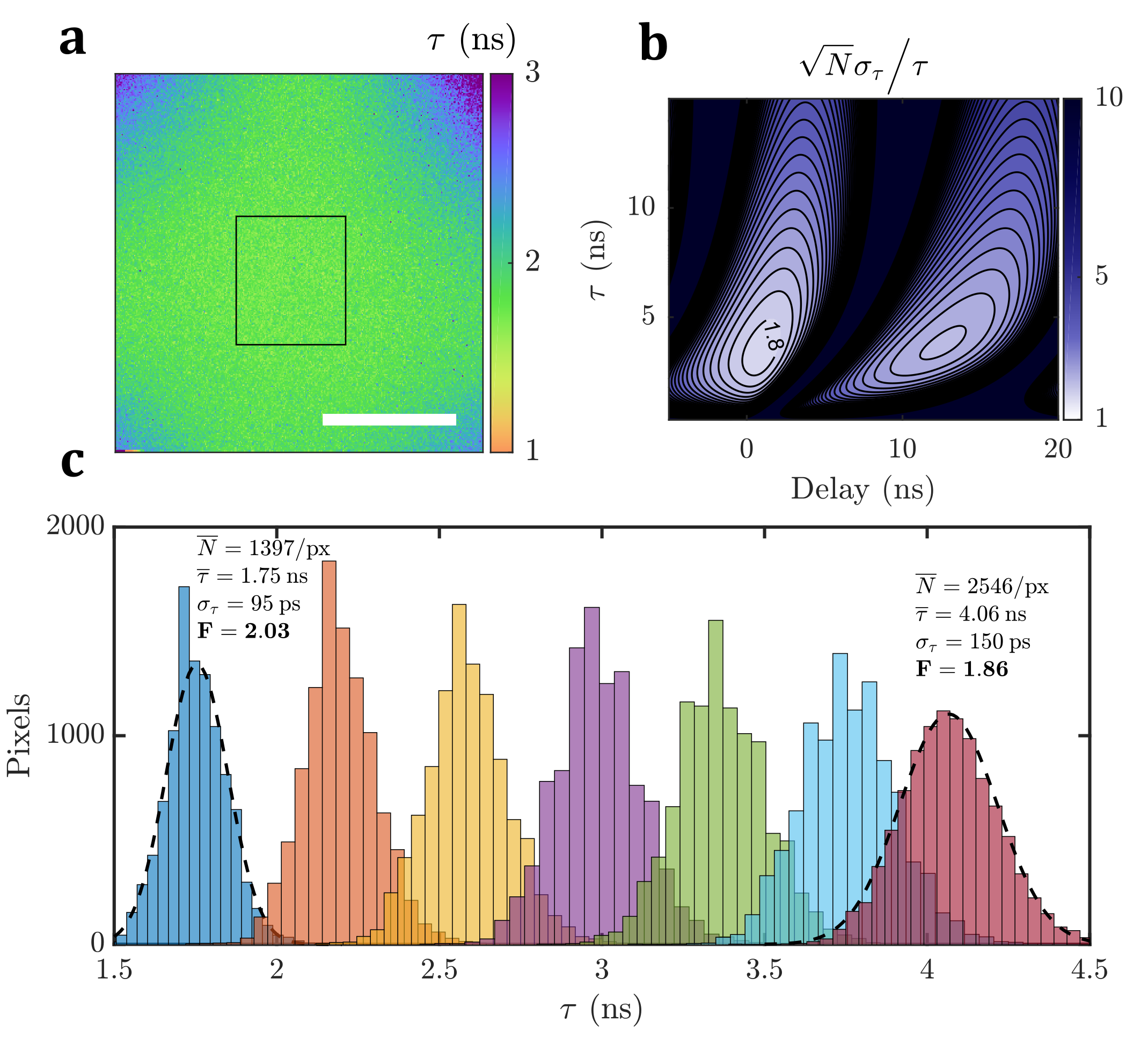} 
\caption{\label{fig:2} {\footnotesize \textbf{Single-frame FLIM estimation.} (a) Wide-field FLIM image of rhodamine B solution acquired in a single exposure near zero time delay (scale bar 50 $\mu$m). (b) Theoretical estimation sensitivity calculated using the  measured gating function (see Methods). (c) Experimental characterization of sensitivity using rhodamine B and rhodamine 6G solutions. Molar concentrations B:6G are in ratios (left to right) 1:0, 5:1, 2:1, 1:1, 1:2, 1:5, and 0:1. Pixel histograms are plotted for the 100x100 pixel ROI indicated in (a). From the mean signal per pixel and the fit histogram width, we can infer a lifetime sensitivity of $F = 2.03$ at $1.8$ ns ($95$ picoseconds at $1397$ photons / pixel) and $F = 1.86$ at $4.1$ ns lifetime ($150$ ps at $2546$ photons / pixel). This is consistent with the expected sensitivity in (b) near zero delay that is directly calculated from the measured gating function.
}}
\end{figure}
\noindent
\textbf{Rapid wide-field FLIM for bright samples.}  Resonant EO-FLIM enables high-throughput FLIM in wide-field with millsecond exposure times. This capability is desirable for imaging fast cellular signaling, particularly for voltage imaging in neuroscience applications\cite{Brinks2015Two-PhotonVoltage,Raspe,Lazzari-Dean2019OpticalImaging}. Using the single frame estimation method, wide-field acquisition on two samples is demonstrated in Fig. \ref{fig:3}. A mouse kidney section labelled with Alexa 488 and Alexa 568 is imaged in a 20 millisecond exposure, allowing the two labels to be differentiated by lifetime within a single spectral channel. A standard FLIM characterization sample, \textit{Convallaria majalis} stained with acridine orange, is acquired in 1 millisecond total exposure and $>10^8$ photons per frame. This acquisition speed is 3 orders of magnitude faster than recent technical demonstrations on the same sample at similar SNR \cite{Popleteeva2015FastImaging,Zickus2020FluorescenceEstimation}, and it corresponds to 5 orders of magnitude higher throughput than conventional TC-SPC confocal acquisition at 2 MHz effective count rate. We note that this throughput could also have been achieved at lower repetition rates on densely labelled samples. However, MHz rates are advantageous for many applications since they are compatible with standard mode-locked lasers and reduce peak power and photo-damage to the sample \cite{Debarre2014MitigatingRange}.\\

\noindent
\textbf{Lifetime sensitivity. }  The lifetime estimation sensitivity of resonant EO-FLIM depends on the slope of the modulation function, the time (phase) delay used, and the number of photons collected. For optimal sensitivity a phase near 0 or 90 degrees is chosen. The theoretical lifetime sensitivity may be directly inferred from the slope of the estimation function and the shot-noise in intensity bins G and U (see Methods). The photon-normalized sensitivity for lifetime estimation is quantified by $F = \sqrt{N} \sigma_\tau/\tau$, where $N$ is the number of detected photons and $F = 1$ corresponds to the photon shot-noise limit \cite{Kollnerr1992, Gerritsen2002FluorescenceResolution,Esposito2007,Turgeman2013PhotonSystems,Zhang2016InvestigationMicroscopy}. With the modulation function shown in Fig. \ref{fig:1}(b) a temporal sensitivity of 2.4 ns$/ \sqrt{N}$ can be achieved in measuring the PC phase --- \textit{i.e.}, using pulse time as the unknown parameter. For estimating fluorescence lifetime the modulation function yields a peak theoretical lifetime sensitivity of $F = 1.7$ with sensitivity depending on the measured lifetime and on the phase delay chosen as plotted in Fig. \ref{fig:2}(b). Estimation sensitivity is experimentally verified in Fig. \ref{fig:2} from pixel lifetime histograms. An experimental single pixel lifetime sensitivity of F = 1.86 is achieved which corresponds to $\sigma_\tau = 150$ ps  at 4.1 ns lifetime and an average signal of $N=2550$ photons per pixel. Sensitivity per pixel may be optimized for a given application and lifetime range by adjusting the number of detected photons (exposure time) and using an optimal RF drive frequency. The EO-FLIM method combines high photon efficiency and throughput, and it approaches (within a factor of two) theoretical limits for FLIM imaging efficiency. \\

Resonant EO-FLIM allows wide-field lifetime imaging at high frame rates and with the sensitivity required for single-molecule applications. We demonstrate initial capabilities for wide-field FLIM-TIRF, single-molecule super-resolution, imaging of fast single-molecule dynamics, and FLIM with high photon throughput. We expect that EO-FLIM and resonant electro-optic modulation will enable time-domain contrast in a variety of wide-field microscopy applications. 

Compared to our previous demonstration \cite{Bowman2019Electro-OpticMicroscopy}, resonant EO-FLIM improves the sample excitation rate by 8000x, allowing optimal excitation of low light samples such as single molecules. Resonant drive also overcomes practical limitations of high-voltage pulse electronics and piezoelectric resonances of EO crystals that are encountered when operating at high repetition rates. By capturing gated and ungated intensity simultaneously, EO-FLIM is self-normalizing for intensity variations that occur frame-to-frame. This is critical for FLIM imaging of fast dynamics. Rapid FLIM has been previously achieved in both time and frequency domains \cite{Schneider1998RapidApplications,Elson2004,Raspe, Sparks2017}, but wide-field detectors have either involved high photon loss or high pixel noise. Modulated CCD and CMOS sensors for frequency-domain FLIM have $> 400$ e/s dark current and $> 30$ e read noise. Similarly, methods using gated intensifiers have low QE and often involve multi-phase or multi-delay sampling that is performed in sequential exposures since the ungated light is not captured. Sequential exposures are not suitable for imaging  dynamics since intensity variations affect the measured lifetime. This problem can be overcome by dividing the GOI into multiple separately gated segments to simultaneously capture several delays\cite{Elson2004}, however this has the trade-off of further lowering photon efficiency.

While EO-FLIM enables a powerful combination of photon efficiency, high throughput, and low sensor noise, a few trade-offs are necessary in order to implement the technique. The camera sensor’s area must be divided to accommodate two or four image copies, or multiple sensors must be used. The range of lifetimes that can be efficiently estimated is determined by resonant frequency, which may need to be changed depending on the sample. Finally, resonant drive at high voltage can also cause significant heating of the EO crystals and tank circuit. Thermal effects  limit our run time to 40 seconds in the current configuration, but future designs will incorporate improved thermal management to allow continuous operation.

Above we focused on rapid single-frame estimation of mono-exponential fluorescent decays with one unknown lifetime parameter for single-molecule applications. Multi-exponential analysis is also a valuable approach in many FLIM experiments and will be compatible with EO-FLIM. Multi-exponential information may be acquired by imaging with separate delays on PC1 and PC2, by using series configurations of PC modulators, or by sequentially imaging at multiple phase values (similar to the dye traces shown in Supplementary Fig. 3). Phasor-based analysis techniques may prove fruitful for rapid multi-exponential imaging \cite{Digman2008TheAnalysis}. We expect electro-optic imaging methods will be feasible at frequencies into the 100's of MHz. Future system improvements will also allow increased field of view through optimized PC crystal geometries.

\noindent
\subsection*{Conclusions}
\noindent
EO-FLIM enables all-optical image processing that is entirely independent from the sensor used. This makes it compatible with any wide-field technique including light-sheet \cite{Greger2011Three-dimensionalRatio,Weber2015MonitoringMicroscopy,Mitchell2017FunctionalMicroscopy,Funane2018SelectiveSamples,Hirvonen2020LightsheetCounting}, TIRF illumination \cite{Blandin2009Time-gatedSource,Giraud2009FluorescenceMicroarrays,Giraud2010FluorescenceImager,Hirvonen2016PicosecondDetector} and spinning-disc confocal microscopy. It also makes EO-FLIM naturally suited to applications where certain detector characteristics are desired or where multiple detection dimensions are combined \cite{Eggeling2001DataDetection, Prummer2004MultiparameterAnalytics,Squires2017DirectC-phycocyanin,Esposito2019EnhancingMicroscopy,Chacko2021HyperdimensionalFiber}. Applications in high-throughput single-molecule imaging and light-sheet microscopy\cite{Greger2011Three-dimensionalRatio,Weber2015MonitoringMicroscopy,Mitchell2017FunctionalMicroscopy,Funane2018SelectiveSamples,Hirvonen2020LightsheetCounting} appear particularly promising. Further, we anticipate that next-generation camera sensors will allow impressive capabilities when combined with nanosecond EO modulation. Examples could include photon number resolving cameras \cite{Tiffenberg2017Single-ElectronCCD, Ma2017Photon-number-resolvingGain} and high well-capacity cameras for improving shot-noise limited measurements \cite{Hosseini2016PushingMicroscopy}. We note that EO imaging has also been applied recently to kilohertz image modulation \cite{Panigrahi2020AnFrequency, Jouchet2021NanometricExcitation} and beam profiling applications\cite{Cao2020OpticalDetectors}.

Wide-field FLIM may provide valuable information and contrast mechanisms for many applications in microscopy. These include improved label multiplexing \cite{Niehorster2016a,Valm2016MultiplexedLabels,Valm2017}, lifetime FRET detection \cite{Sorokina2009FluorescentInitiation, Poland2015AImaging, Levitt2020QuantitativeFLIM}, distance measurement by metal-induced energy transfer \cite{Cade2010ALifetime,Cade2013Plasmon-assistedImaging,Chizhik2014Metal-inducedNanoscopy,Gregor2019Metal-inducedTransfer}, lifetime imaging of dynamic processes and signaling \cite{Brinks2015Two-PhotonVoltage,Raspe,Lazzari-Dean2019OpticalImaging}, and single-molecule spectroscopy \cite{Eggeling2001DataDetection,Prummer2004MultiparameterAnalytics, Squires2017DirectC-phycocyanin, Oleksiievets2020Wide-FieldMolecules}. FLIM contrast in tissue may be obtained from endogenous cofactors like NAD(P)H to enable imaging of cellular metabolism and label-free clinical diagnostics \cite{Datta2020FluorescenceApplications, Konig2020Review:Tomography}. Multi-dimensional imaging will also allow improved information content in biophysical measurements that have limited photons\cite{Watkins2004InformationMeasurements, Esposito2013MaximizingMicroscopy}. As an initial example, here we combine lifetime and spectral FRET detection channels and also show simultaneous polarization anisotropy and lifetime imaging. We also anticipate that lifetime information may be used to improve the spatial and chemical separability of fluorescent probes in super-resolution microscopy \cite{Dong2016, Zhang2015Ultrahigh-throughputMicroscopy}.

Beyond FLIM, resonant EO imaging will allow wide-field lock-in detection at high modulation frequencies. This may prove useful for polarization-modulated microscopy techniques to image molecular orientations \cite{Ha1996SingleModulation,Backer2016EnhancedMeasurements,Chandler2020Spatio-angularImaging} and also for label-free techniques that rely on signals detected by modulation transfer such as photothermal \cite{Boyer2002PhotothermalScatterers, Bai2019UltrafastAbsorption}, stimulated emission \cite{Min2009}, and stimulated Raman microscopy \cite{Saar2010Video-rateScattering}. High frequency lock-in detection improves rejection of low frequency 1/f noise sources for optical metrology.  We also foresee applications for quantum imaging, imaging in scattering media, and wide-field LIDAR. 

\noindent

\bigskip
\noindent\textbf{\large Methods}

\bigskip
\noindent\footnotesize{\textbf{Optical system.} 
A Zeiss Axiovert S100TV microscope is used with either a 60x 1.49NA Olympus APO N 60X TIRF objective (Figs. 1-4) or a 20x 0.8NA Zeiss PlanApo objective (Figs. 5 and 6). The objective is mounted on a piezo stage (Physik Instrumente P-721) to allow fine focus control. The bottom output port of the microscope is routed to the detection optics shown in Fig.  \ref{fig:1}(a). These comprise a 4f relay that creates a Fourier plane near each of the PC locations. Polarizing beamsplitter cubes (Thorlabs PBS251) split polarizations before and after each PC. Images are then combined by using a right-angle prism mirror (Thorlabs MRAK25-P01) and imaged onto an Andor Neo5.5 sCMOS camera. Excitation is provided by a 78 MHz mode-locked supercontinuum laser with internal pulse-picker to halve the repetition rate (NKT SuperK EXW-12). The supercontinuum source is filtered with a 532 nm short-pass (Semrock BSP01-532R) and a colored glass filter (Thorlabs FGS900). The excitation band is then further restricted by a 40 nm bandpass at 520 nm (Thorlabs FBH520-40). For fluorescence detection, a 532 nm dichroic mirror (Semrock di03-R532-T1) and 532 nm long-pass filter (Semrock BLP01-532R) are used for all data presented here. A translation stage allows switching between epi and TIRF illumination modes. For FRET detection, a shortpass filter is added to one EO-FLIM detection path (Thorlabs FESH600) to image  the donor channel, while a long-pass dichroic is used to image the acceptor emission in the second EO-FLIM path (Semrock FF660-Di02-25x36).

The two PCs are modified 10 mm aperture longitudinal field dual-crystal KD*P units with 40 mm length  (Lasermetrics 1072). The crystal coatings are index matched to and immersed in fluorocarbon oil, so these units are easily modified to allow for flowing coolant. Fluid adaptor ports were attached at the two window seals with viton gaskets, and a recirculating fluorocarbon-compatible chiller (solid state cooling systems ThermoCube FC ) and flow control manifold was used to provide cooling (for details see Supplementary Fig. 1). 3M Fluorinert FC-40 fluid was used for the optical cooling loop. Cooling lines were nylon tubing, and connections were made with swagelok where possible. Thread joints were sealed with Leak Lock Blue (Highside Chemicals). \\
}

\noindent\footnotesize{\textbf{RF system.} 
The RF drive is synchronized to the supercontinuum laser using a Novatech 409B direct digital synthesizer (DDS). The sync pulses from the laser are low pass filtered (Mini-Circuits BLP-90) and used as the external clock input to the DDS. RF is generated on two channels at 19.5 MHz, which is half the laser repetition frequency. RF phase can be set arbitrarily for each PC channel and may be updated at 40 Hz during image acquisition. RF switches (Mini-Circuits ZYSW-2-50DR) and variable attenuators (Mini-Circuits ZAS-3+) allow control of the RF level. Two class-A amplifiers (ENI 325LA) are used to drive the tank circuits. T-network antenna tuners \cite{Terman1943RadioHandbook} are used for impedance matching the tank circuit to 50 ohm drive and also for monitoring SWR (Palstar AT2K). LMR-400 coaxial cables with type N connectors link the matching networks to the tank inductor boxes that sit on top of each PC. A series-driven tank circuit is used. Tank inductors are wound from 7 turns of 3/16" copper tubing with a 2 inch inner diameter (see Supplementary Fig. 1).  A trim capacitor (1-10 pF, Sprague-Goodman SGNMNC2106) allows fine adjustment of the resonant frequency to 19.5 MHz. Tuning and impedance matching is performed while monitoring S11 on a vector network analyzer (HP 4395A / 87511A). Care must be taken to ensure good grounding and RF shielding in order to minimize interference and spurious emission.\\
}

\noindent\footnotesize{\textbf{Sample details and preparation.} 
Dye standards in Supplementary Fig. 3 were prepared in either water  (erythrosin B - solvent red 140, rhodamine B, rhodamine 6G ) or methanol (rose bengal and rubrene). Dyes were sourced from Sigma Aldrich. For the rhodamine B and 6G mixtures in Fig.  \ref{fig:2}, 0.9 mM concentrations were used. In Fig.  \ref{fig:3}, commercially available slides are imaged. The mouse kidney section is ThermoFisher FluoCells slide $\#$3 and the acridine orange stained \textit{Convallaria majalis} was sourced from MSMedia but is widely available. In Fig.  \ref{fig:4}, PMMA films are made by spin coating a 1$\%$ wt solution of PMMA dissolved in toluene at 3000 rpm. Alexa 532, Alexa 546 (ThermoFisher) and  quantum dots (CdSe/ZnS,  540 nm emission, Sigma Aldrich 748056) are diluted in the PMMA toluene solution to achieve single-molecule concentrations. Coverslips (22 mm, Marienfeld $\#$1.5H) are cleaned before spin coating by sonicating for 20 minutes in a 2$\%$ solution of Hellmanex III (Sigma Aldrich), rinsing in DI water, and then drying with compressed air. DNA origami samples are obtained from Gattaquant. For DNA-PAINT super-resolution in Fig.  \ref{fig:5}, Gattaquant's standard PAINT 80G sample is used which has three binding sites separated by 80 nm. For two-dye PAINT, a custom sample with Cy3B and Atto 565 is used with with four binding sites per origami having 80-80-20 nm spacings. For FRET, Atto 532 donor and Atto 647N acceptor molecules are used on a custom origami structure having 4.1 nm fixed spacing.\\

}

\noindent\footnotesize{\textbf{Thermal considerations.}
In order to minimize image distortions from RF-induced thermal heating of the Pockels cells and surrounding cooling fluid, the RF is only turned on while imaging and we are limited to run times $< 40$ seconds. For single-molecule localization data (Fig.  \ref{fig:5}), four separate 40 second acquisitions of 250 frames each are combined, interspersed by cooling intervals to minimize image distortion. Further, only one PC is operated at a reduced RF power so that localization is performed from a non-modulated image and combined with lifetime estimation from the modulated images. This minimizes image motion and drifts. Since the system is not operated in thermal equilibrium, there are also RF phase drifts in the modulation waveform during long runs. This phase drift is pre-calibrated when measuring the gating functions and it is corrected in the single-molecule localization data. All other data uses a single phase value for shorter acquisitions where the drift effect is negligible. Future designs will avoid immersed cooling and thermal complications by using improved electro-optic configurations.\\
}

\noindent\footnotesize{\textbf{Calibrations.}
Lifetime estimation requires calibration of the gating functions in each channel. These gating functions correspond to the instrument response function in an EO-FLIM measurement.  Particularly important is accounting for any wavelength-dependent gating in the Pockels cell. Note that wavelength dependence is minimized by operating the Pockels cell near $V_\pi$. However, for these demonstrations we are at a somewhat lower voltage depending on applied RF power $\sim 0.7-0.85 \:V_\pi$ at 532 nm (or $\sim 0.65$ $V_\pi$ for the long acquisitions in Fig.  \ref{fig:5}). A series of bandpass filters is applied to characterize how wavelength impacts the gating function in order to account for standard fluorescence Stokes shifts (Supplementary Fig. 7). Wavelength calibration can be accomplished by two methods: (1)  bandpass filter measurements may be used to determine the change in gating function fit parameters expected --- \textit{i.e.}, accounting for the reduced modulation depth at longer wavelengths; (2) A calibration sample may be used and then deconvolved using the known fluorescence lifetime to recover the corrected gating function. For method (1), the gating functions are measured by imaging laser reflection from frosted glass. This is done by replacing the detection filter set with a cut glass slide to act as beamsplitter (a separate slot in our filter slider). We found that method (2) is the most convenient and used an orange fluorescent plastic slide (Thorlabs FSK4) with a 5 ns lifetime.

Due to the longitudinal KD*P Pockels cells used here, the field of view for imaging is limited to $\sim$150 microns (20x objective) or $\sim$30 microns (60x TIRF objective). This happens due to off-axis birefringence effects in these crystals \cite{Bowman2019Electro-OpticMicroscopy}. This is not a fundamental limitation of the technique, and larger fields of view will be achieved in the future by using optimized electro-optic crystal geometries. The usable field of view may also be increased and spatial lifetime variations minimized by calibrating the gating function locally. This additional calibration step is used for the super-resolution data in Fig.  \ref{fig:5}. The image is divided into local blocks of pixels with gating function calculated separately for each block to minimize variations across the field of view.\\
}

\noindent\footnotesize{\textbf{Data analysis.}
Image registration is performed by applying affine transformations to align the output images --- \textit{i.e.}, pixels in the gated image are aligned to pixels in the ungated image. The required transformation is determined by imaging a grid micrometer slide (Electron Microscopy Sciences S29) in bright field. This achieves the pixel-to-pixel mapping needed for lifetime estimation.

The Pockels cell applies a birefringent phase shift $\delta = 2 \pi r_{63} V n_\mathrm{o}^3/\lambda$, determined by applied voltage V, ordinary index of refraction $n_\mathrm{o}$, and the longitudinal electro-optic coefficient $r_{63}$. After the second beamsplitter, transmission in the gated and ungated channels is $g(t) = \sin^2[\delta(t)/2]$ and $u(t) = \cos^2[\delta(t)/2]$. In practice g(t) and u(t) are measured directly as the instrument response function.

To calculate lifetime from G/U intensity ratio, the gating functions g and u are first fit using four parameters each
\begin{align*}
g(t) &= p_1\sin^2[p_2\frac{\pi}{2}\sin(\omega t - p_3)]+p_4\\
u(t) &= p'_1\cos^2[p'_2\frac{\pi}{2}\sin(\omega t - p'_3)]+p'_4.
\end{align*}
These four parameters allow accurate fitting of all experimental gating curves. When using the lifetime reference slide, we instead fit the convolution of the above functions with a known 5 ns decay. The four fit parameters $p, p'$ are wavelength corrected if necessary and then convolved with mono-exponential decays at the chosen phase delay $\phi$ in order to generate the ratio to lifetime lookup table (Supplementary Fig. 4) Phase delays should be chosen which provide high estimation sensitivity [Fig.  \ref{fig:2}(b)] and also a monotonic function in the lifetime range of interest. Phase points near 0 and 90 degrees (on the 19.5 MHz RF waveform) give highest theoretical sensitivity. Phase is related to time delay by $\phi_0 = \omega t$. Here all data is taken at 0 degrees except for FRET in Fig.  \ref{fig:6}(b) taken at -15 degrees, and wide-field FLIM in Fig. \ref{fig:3} taken near 90 deg. In cases where there is a high background, such as our FRET demonstration, it can be advantageous to choose a phase such that more intensity is captured in the gated channel.

Lifetimes and fit curves in Supplementary Fig. 3 are determined by least squares fitting --- \textit{i.e.}, fitting the convolution of the wavelength-corrected gating function with a mono-exponential decay. To determine rotational correlation times in Supplementary Fig.  3(b), polarized excitation is used and the normalized intensity difference in the two gated channels, $(G_\parallel - G_\perp)/I$, is plotted. For fitting, the gating functions are convolved with mono-exponential lifetime and fluorescence anisotropy intensity functions for each detection channel
\begin{align*}
I_\perp(t) &= \frac{I_0}{3} e^{-t/\tau}(1-r_0 e^{-t/\tau_\theta})\\
I_\parallel(t) &= \frac{I_0}{3} e^{-t/\tau}(1+2r_0 e^{-t/\tau_\theta}).
\end{align*}
Here $I_{\parallel}+2I_{\perp} $ gives the total fluorescence intensity emitted in all directions, and we set initial anisotropy $r_0$ = 0.35. This makes fitting simpler due to DC offsets caused by different focus settings in the solution samples and because $r_0$ has little effect on the shape of the traces. In principle, $r_0$ may be determined in addition to $\tau$ and $\tau_\theta$ from the measured traces. Transmission differences in each PC path (typically accounted for as the G-factor) also affect the shape of the measured curves and are included when convolving with the measured gating functions. 

Lifetime estimation sensitivity may be directly calculated from the measured gating functions for the case of a single unknown parameter by error propagation. Here we are interested in recovering lifetime $\tau$ from the intensity ratio R = G/U. For a shot-noise limited measurement this gives ratio error
\begin{equation*}\label{eq1}
    \sigma_{\tau} = \bigg{|}\frac{\partial \tau}{\partial R}\bigg{|}\: R \sqrt{\frac{1}{G}+\frac{1}{U}}.
\end{equation*}
\noindent
This ratio error is converted to lifetime error using the lookup table.

To estimate single-molecule lifetimes, the molecule is first localized by Gaussian fitting and a square ROI of 6x6 pixels is summed to determine the molecule's intensity. Proper background subtraction is critical, as it affects the measured ratio and can easily bias lifetime measurements. Background is determined by applying Gaussian filtering to the image ($\sigma = 20$ pixels). For the super-resolution images (Fig. \ref{fig:5}) the bright molecules are first removed before Gaussian filtering ($\sigma = 10$ pixels) by using a morphological opening filter with a 5 pixel disc structural element.  

For single-molecule localization, we combine the lifetime image (gated and ungated frames) with an intensity image from the other detection path that is not modulated. This minimizes the effect of fluid-induced image shifts. Single-molecule localization images are analyzed using the ThunderSTORM imageJ plugin \cite{Ovesny2014ThunderSTORM:Imaging} to generate a super-resolution image. The localization coordinates are then used to compute the intensity ratio and lifetime estimates from the gated and ungated image frames. 

In order to determine lifetime error bars (standard error) for single molecules, the error in the measured intensity ratio is calculated by combining background error, read noise, and shot-noise terms as:

\begin{equation*}
\sigma_R = \frac{G}{U}\sqrt{\frac{1}{G} + \frac{1}{U }+ N_p(\frac{\sigma_\mathrm{G}^2+\sigma_{RN}^2}{G^2}+\frac{\sigma_{\mathrm{U}}^2+\sigma_{RN}^2}{U^2})}\:.
\end{equation*}where $N_p$ is the number of pixels summed, read noise $\sigma_{RN}$ is measured experimentally from dark frames and is scaled according to the number of frames used per lifetime estimate, and Poisson background error $\sigma_\mathrm{G}$ and $\sigma_\mathrm{U}$ are determined from mean background intensities in the frames (again summing over frames used in each lifetime estimate).  

When performing multi-parameter imaging a polarizer is inserted into the excitation path. Fluorescence polarization $P =$ $I_\parallel - I_\perp$ may be determined after correcting for the transmission efficiency of the two channels, and error bars are calculated similar to above. For multi-parameter imaging of single-molecule binding in Fig.  \ref{fig:6}, excitation must be polarized such that s-polarized light is used for TIRF illumination.
}

\bigskip


\noindent\textbf{Notes:}\\
\noindent Authors are inventors on PCT/US2019/062640 and US17/153438.\\\\
\noindent{The data that support the findings of this study are available from the corresponding author upon reasonable request.}\\\\

\noindent\textbf{\large Author Contributions}
AB conceived the technique and performed the experiments. AB and MK prepared the manuscript.\\

\noindent\textbf{\large Acknowledgments}
This research was funded by the Gordon and Betty Moore Foundation. A.B. acknowledges support from the Stanford Graduate Fellowship and from the National Science Foundation
Graduate Research Fellowship Program under grant 1656518.

\bibliography{references}

\clearpage
\renewcommand{\figurename}{Supplementary Figure}
\setcounter{figure}{0}

\begin{figure*}[h!]
\centering
  \includegraphics[width=\textwidth]{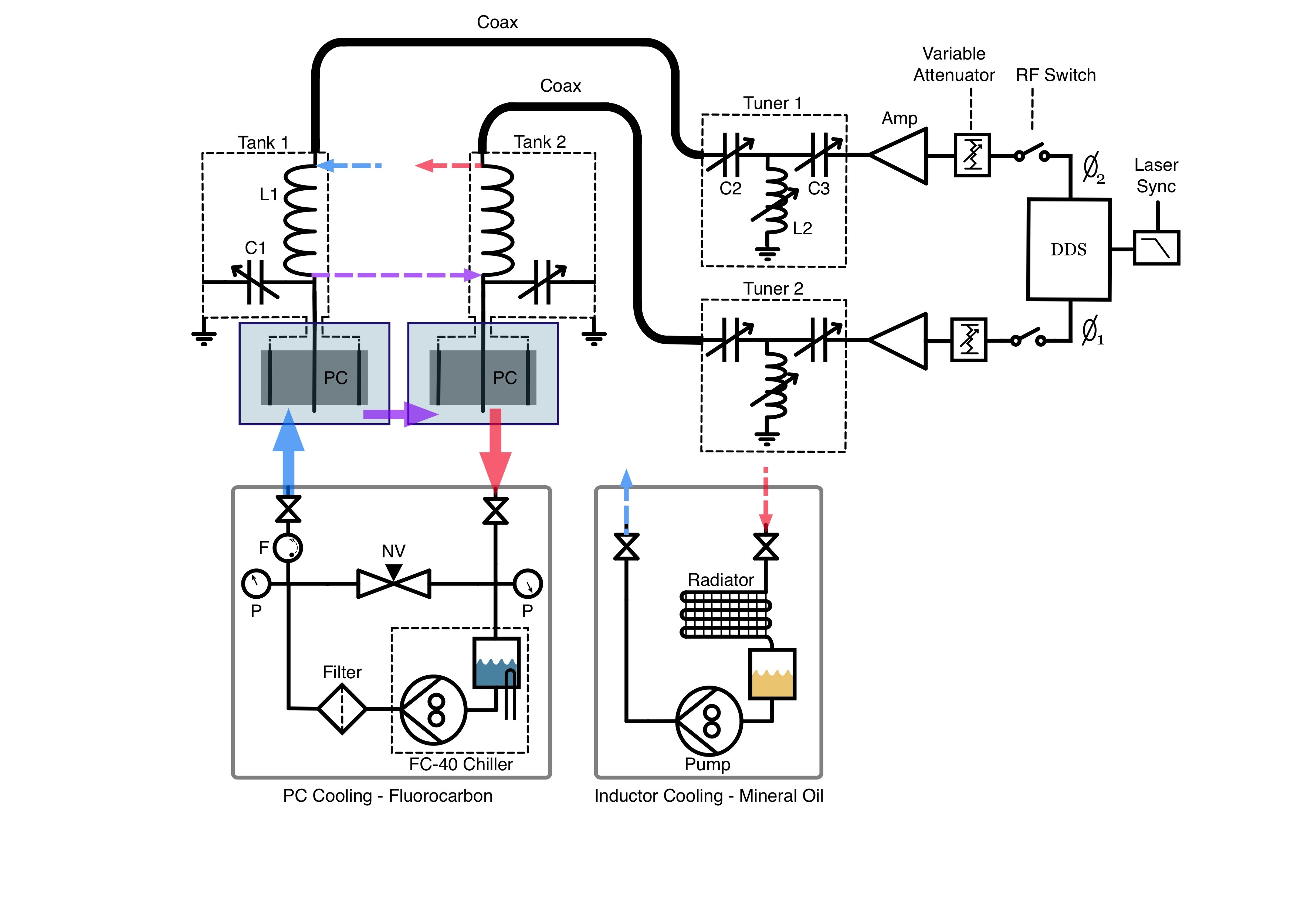} 
  \vspace{-5.0mm}
\caption{\label{fig:s1} {\footnotesize \textbf{RF system schematic.} Sync pulses from the laser are low-pass filtered and used as clock input to a direct digital synthesizer. RF switches turn amplifiers ON/OFF during data acquisition and variable attenuators are used to control the drive amplitude. Two class-A amplifiers give 25-40W of RF power to drive the PCs at 19.5 MHz. T-network antenna tuners are used to impedance match the series tank circuits. The Pockels cells are liquid cooled with FC-40 using a fluorocarbon-compatible gear pump chiller. A 10 micron particulate filter is included since the coolant flows over optical surfaces. Flow rate is controlled by using a needle valve bypass with both pressure and flow rate indicators. A second passive cooling circuit using mineral oil was constructed for the tank inductors using a small gear pump and components from a liquid-cooled computer kit. Though not strictly necessary, this loop helps to increase thermal stability of the tank inductors. See Methods for part numbers.
}}
\end{figure*}

\clearpage
\begin{figure*}[t!]
\centering
  \includegraphics[width=\textwidth]{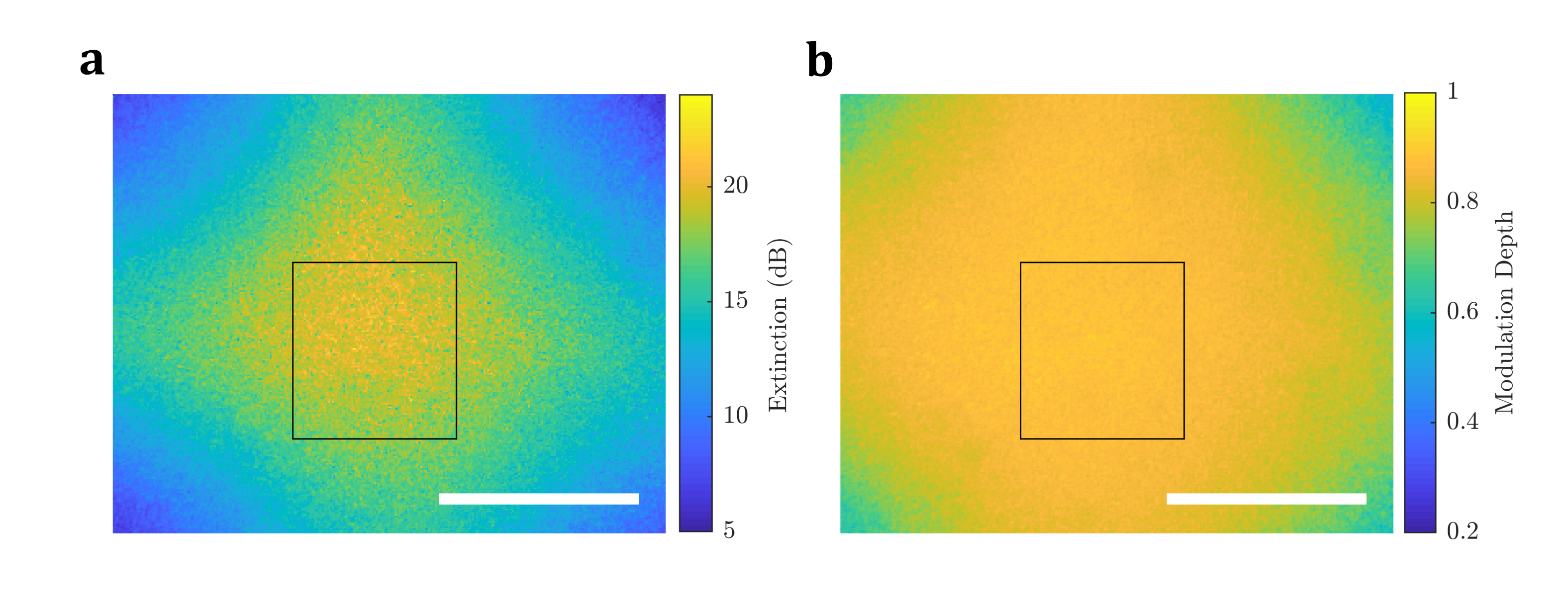} 
  \vspace{-5.0mm}
\caption{\label{fig:s3} {\footnotesize \textbf{Spatial modulation depth and extinction.} Imaging field of view and spatial variation of the gating function can be characterized by imaging laser pulse reflection from of a frosted glass slide (using 20x objective here). (a) Extinction [defined as $10\log_{10}(\text{max/min})$] is calculated for each pixel  in the gated (off at 0 V) channel where max and min refer to extrema of the intensity trace (b) Modulation depth in the ungated channel (on at 0 V) is calculated as (max-min)/(max+min). The ROI corresponds to 100x100 pixels (same as in Fig. 6 -- 41x41 $\mu$m) having average modulation depth of 86$\%$ and extinction of 16.2 dB. Scale bar is 50 $\mu$m. 
}}
\end{figure*}

\clearpage
\twocolumn
\begin{figure*}[t!]
\centering
  \includegraphics[width=\textwidth]{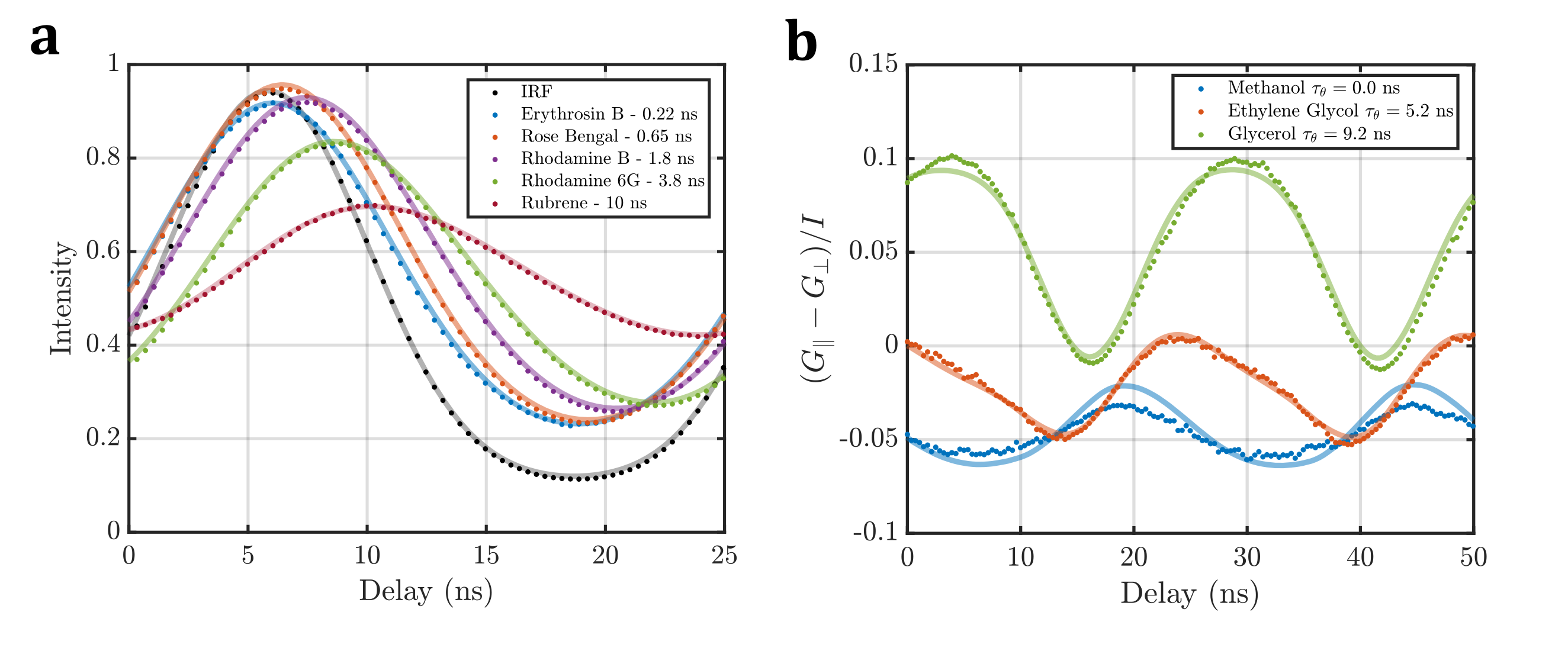} 
  \vspace{-5.0mm}
\caption{\label{fig:s0} {\footnotesize \textbf{Dye solution time traces.}  Time traces are measured by varying the phase of the Pockels cell drive $\phi_0$ relative to the laser pulse train, where trace time $t = \phi_0/\omega$.  (a) Traces are measured on a series of ungated images of various dye solutions measuring from $220$ ps to $10$ ns in lifetime. These are fit to single-exponential decays to determine fluorescence lifetime (best fit plotted with values given). (b) Time-resolved polarization anisotropy is measured by subtracting the two gated channel intensity traces. Rhodamine 6G traces are shown in methanol, ethylene glycol, and glycerol solvents. Fitting these traces allows measurement of the rotational correlation time $\tau_\theta$ in the solvent  (best fit plotted with values given).
}}
\end{figure*}

\clearpage
\begin{figure*}[t!]
\centering
  \includegraphics[width=0.8\textwidth]{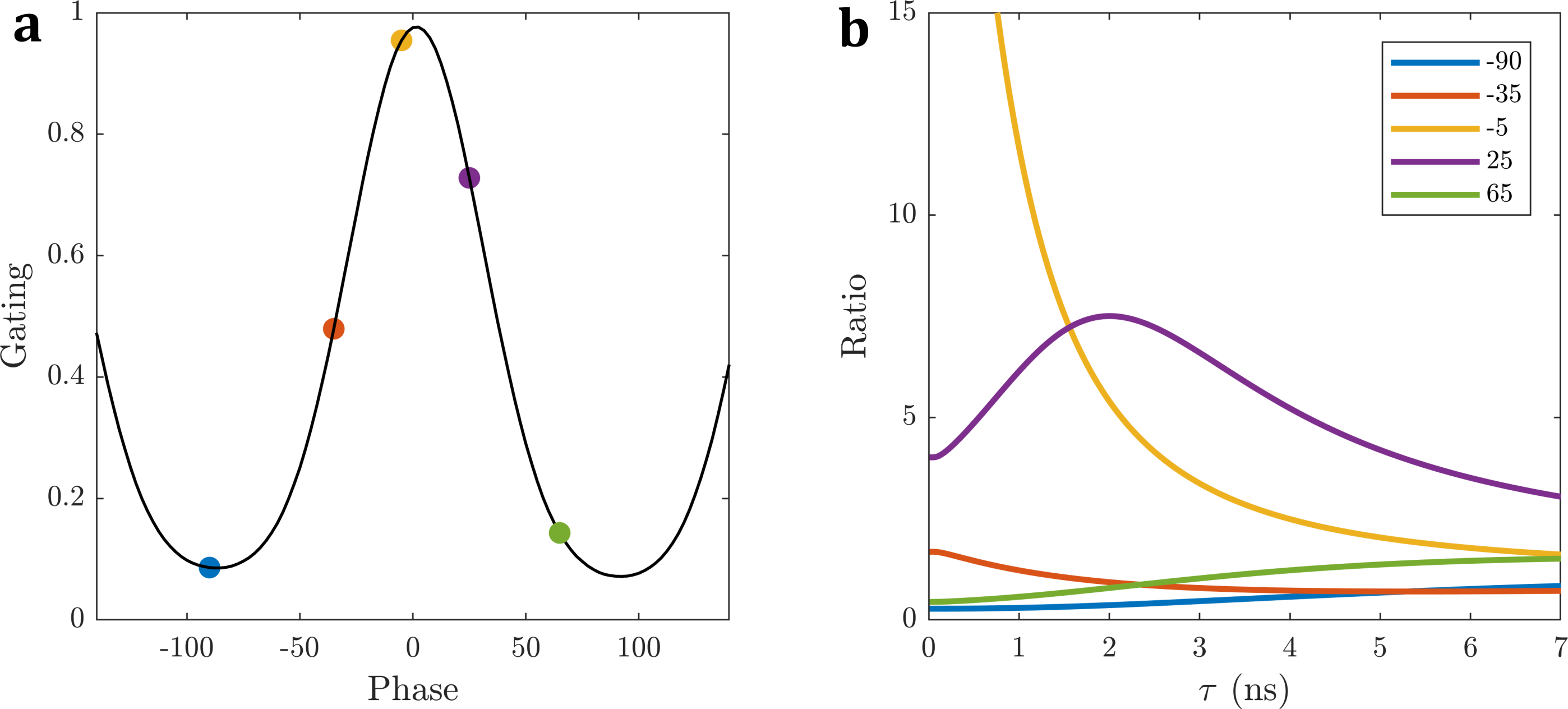} 
  \vspace{-0.0mm}
\caption{\label{fig:s4} {\footnotesize \textbf{Lifetime lookup tables.} Ratio (U/G) to lifetime lookup tables are generated from the experimentally measured gating function. Sensitivity and lifetime dynamic range depend on the chosen phase point (time delay) as shown in Fig. 6(b). Several phase points shown in (a) are used to calculate the lifetime lookup tables plotted in (b). Optimal sensitivity for single-frame lifetime estimation occurs near the peak and trough of this gating function. Phases which give local extrema in the lookup table should also be avoided.
}}
\end{figure*}

\clearpage
\twocolumn
\begin{figure*}[t!]
\centering
  \includegraphics[width=0.65\textwidth]{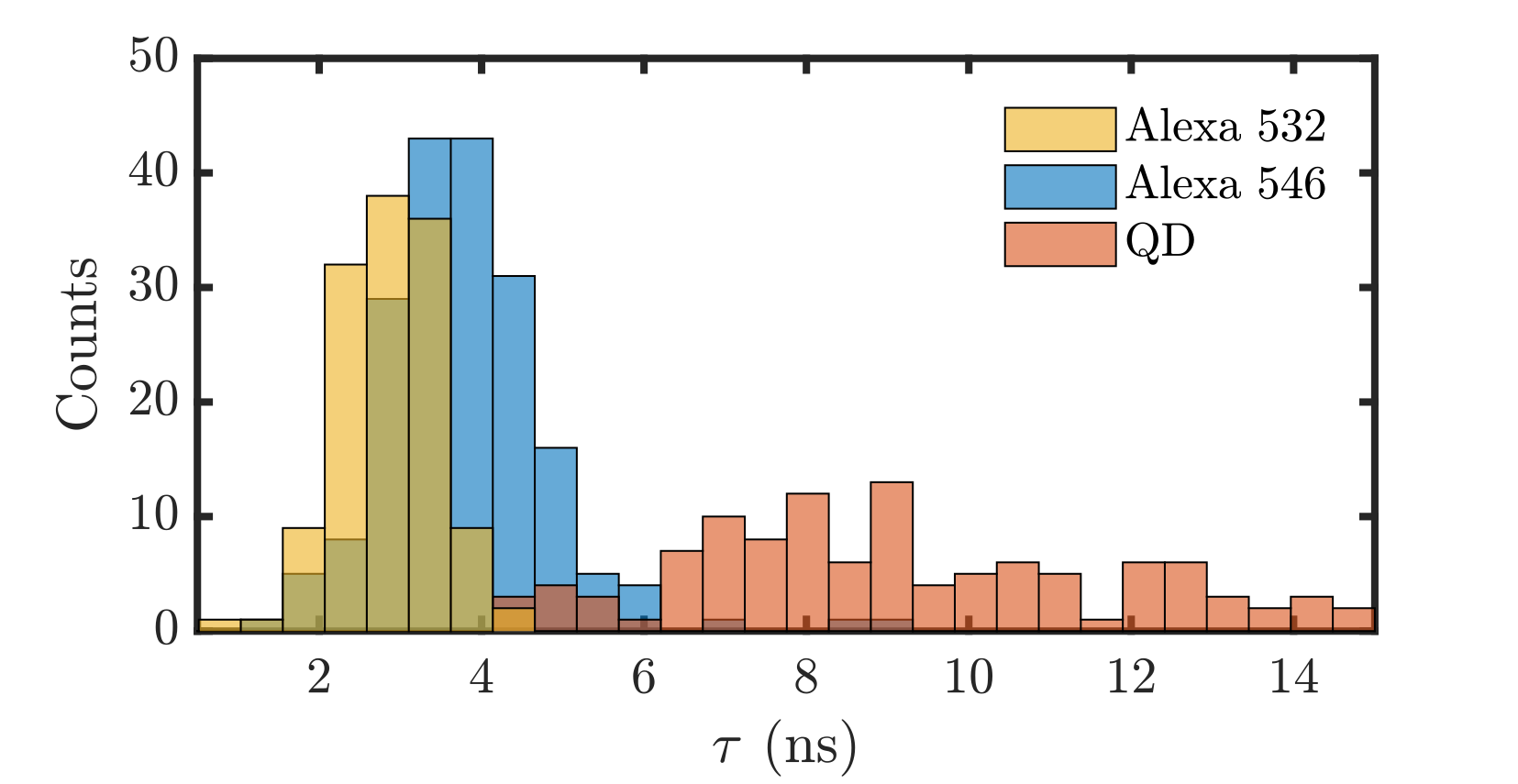} 
  \vspace{3.0mm}
\caption{\label{fig:shist} {\footnotesize \textbf{Single molecule distributions in PMMA.}  Single molecule lifetime distributions are characterized on three separate samples in spin-coated PMMA films. Histograms plot lifetime estimates from single 50 ms (Alexa 532 and QDs) or 100 ms (Alexa 546) frames for molecules giving $> 300$ detected photons in a frame. Spin-coated PMMA films are known to give broad lifetime distributions consistent with what we observe here.\textsuperscript{42,43} 
}}
\end{figure*}

\clearpage
\begin{figure*}[t!]
\centering
  \includegraphics[width=0.8\textwidth]{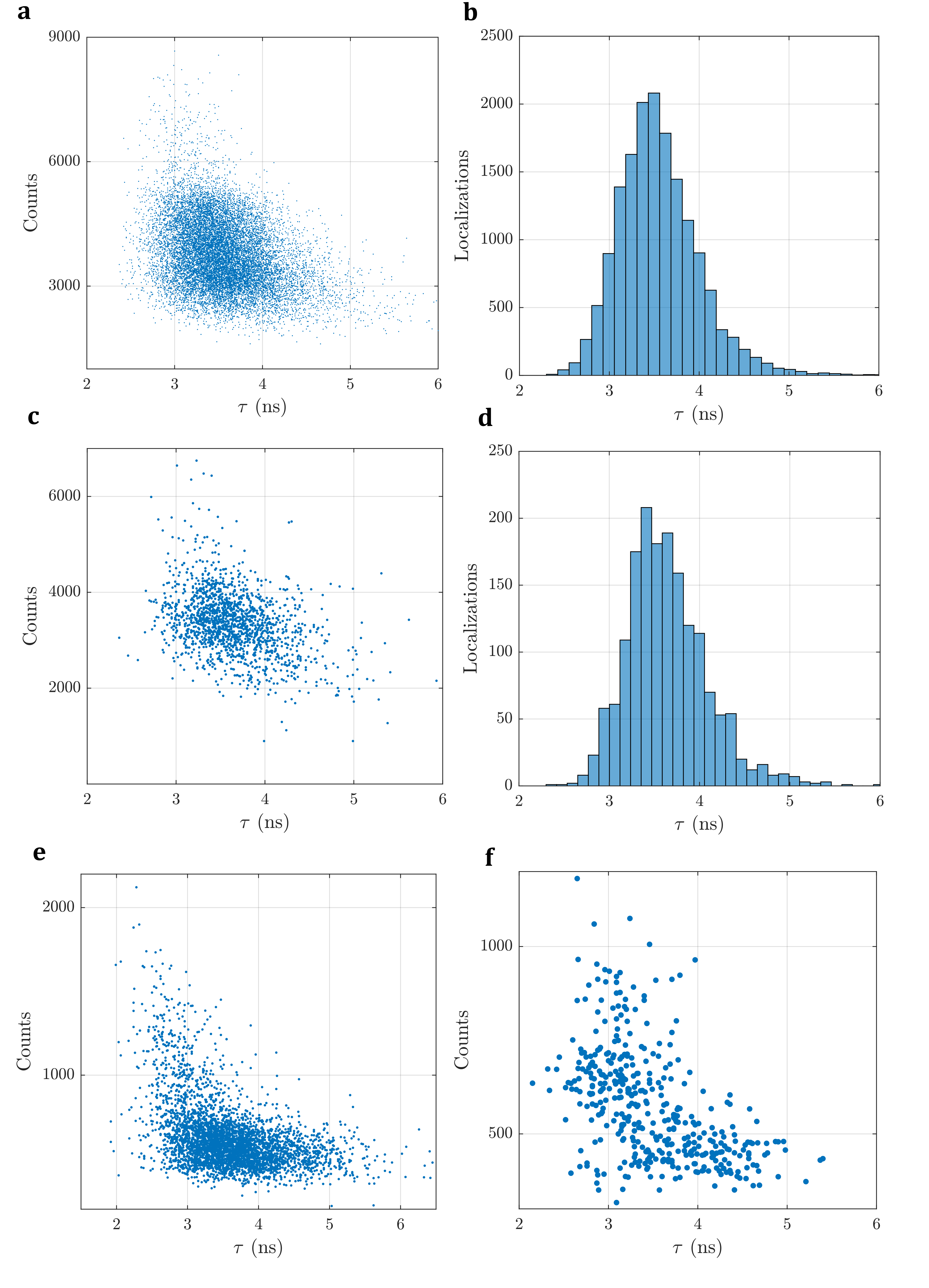} 
  \vspace{0cm}
\caption{\label{fig:s6} {\footnotesize \textbf{Localization distributions for PAINT data.} Each localization in the DNA-PAINT super-resolution data may be characterized by its position, brightness, and fluorescence lifetime. (a-b) Intensity \textit{vs.} lifetime scatter plots and lifetime histograms for a large field of view DNA-PAINT acquisition - using a centered 10 x 15 $\mu$m region of Fig. 3(a) with uniform excitation intensity. (c-d) The same plots for the smaller ROI in Fig. 3(b). (e-f) Intensity \textit{vs.} lifetime scatter plots for the dual dye DNA-PAINT sample corresponding to a large field of view (e), and corresponding to the smaller field (f) shown in Fig. 3(c). Due to the excitation filter used, the Cy3B molecules are brighter.  They tend to cluster at shorter lifetimes and higher intensities than the Atto 565 molecules. Fiducial markers using Atto 565 also give it a higher localization frequency compared to Cy3B in this sample.
}}
\end{figure*}

\clearpage
\begin{figure*}[t!]
\centering
  \includegraphics[width=0.8\textwidth]{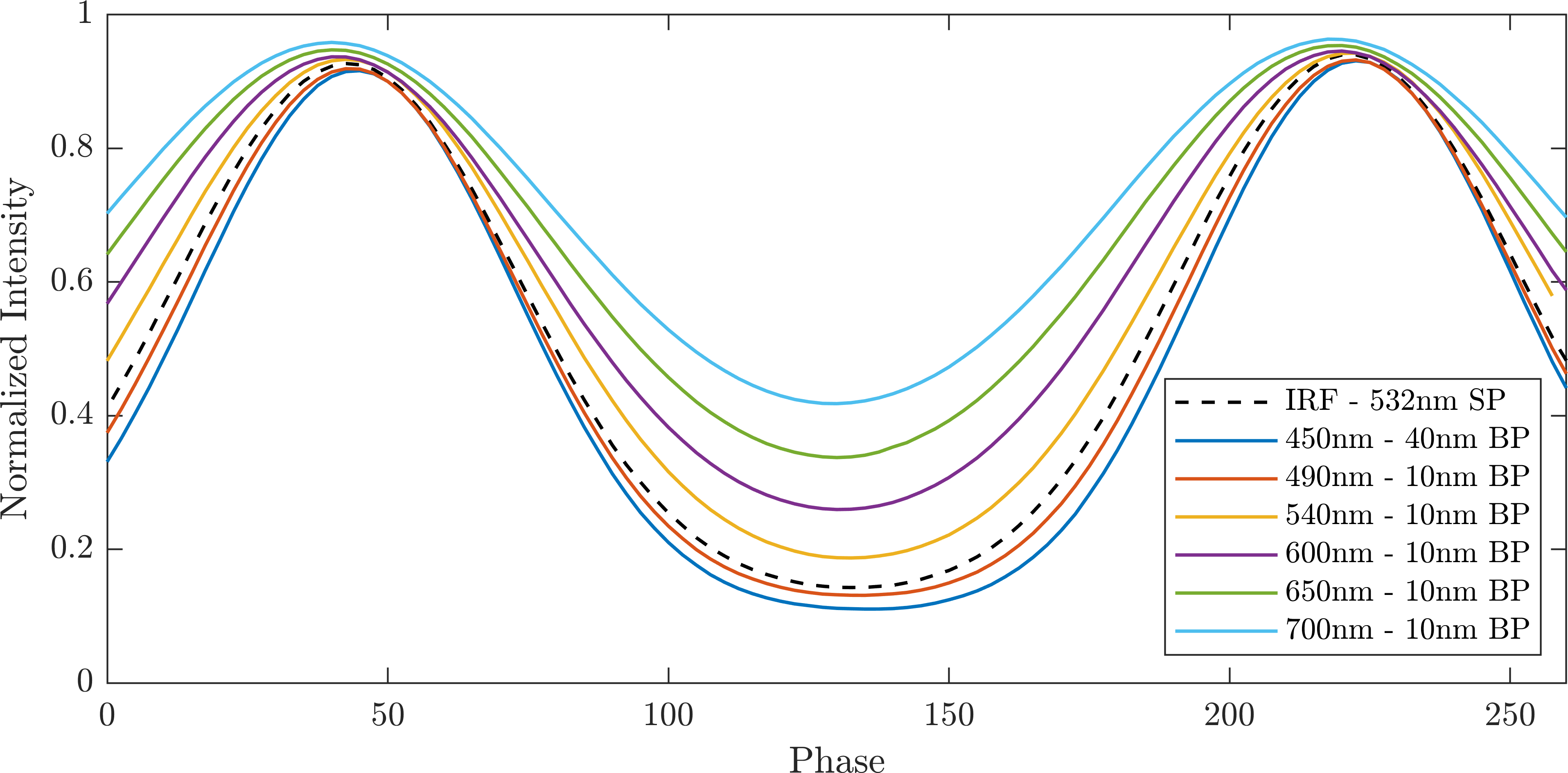} 
  \vspace{5.0mm}
\caption{\label{fig:s5} {\footnotesize \textbf{Wavelength dependence of gating functions.} Wavelength dependence of the gating function is measured on frosted glass using various excitation bands. Near $V_\pi$ there is minimal effect of wavelength changes on the gating functions, but below $V_\pi$ the shift in modulation depth must be taken into account.
}}
\end{figure*}

\clearpage
\onecolumn

\bigskip
\noindent
\small{\textbf{Supplementary Video 1:}  Video of a time trace [Fig. 1(b)] using laser reflection from a frosted glass slide to measure the gating function. Each frame corresponds to a different RF phase on the PCs, with the corresponding time delay displayed for phases between 0 and 360 degrees to show two modulation cycles (scale bar is 100 $\mu$m). Here the video is formed by changing the relative phase frame-to-frame in order to measure the gating function. During EO-FLIM acquisition this modulation is applied to the wide-field image at 39 MHz.

\bigskip
\bigskip

\noindent
\textbf{Supplementary Video 2:} Video demonstrating wide-field single-molecule FLIM on a DNA-PAINT sample in TIRF microscopy (Gattaquant 80G). Lifetime is estimated for each molecule in each 150 ms frame (scale bar 10 $\mu$m).
}

\end{document}